# Studio di un urto anelastico: una proposta per le Scuole Secondarie di II grado nell'ambito del progetto "*Lab2Go*"

Pia Astone[1], Roberto Balaudo[2], Fausto Casaburo[3,1,*], Francesca Cavanna[4], Giulia De Bonis[1], Riccardo Faccini[3,1], Davide Fallara[5], Andrei Grigoruta[2], Giovanni Organtini[3,1], Francesco Piacentini[3,1], Francesco Pennazio[4]

[1] *Istituto Nazionale di Fisica Nucleare (INFN)- Sezione Roma*

[2] *IIS Curie Vittorini-Grugliasco (TO)*

[3] *Sapienza Università di Roma- Dipartimento di Fisica*

[4] *Istituto Nazionale di Fisica Nucleare (INFN)- Sezione Torino*

[5] *Università degli Studi di Torino- Dipartimento di Fisica*

[*] *Corresponding author*

ABSTRACT

When a free falling ping-pong ball collides on a horizontal surface, it loses kinetic energy. The ratio between the height reached by the ball after the collision and the initial height is called restitution coefficient. A method to measure it by using a home-made cathetometer was proposed during the *Olimpiadi di Fisica 2018.* In this paper we show how to measure it also by using the *PhyPhox* app and *Arduino* board.

**Introduzione**

È esperienza comune che, se lasciamo cadere una pallina al suolo da un'altezza $h_0$, essa rimbalzerà raggiungendo un'altezza $h_1<h_0$. Durante la collisione col suolo, la pallina emette un suono e si genera attrito tra le molecole durante la fase di compressione producendo un aumento di temperatura; inoltre, durante la sua risalita, il suo moto è decelerato dalla resistenza dell'aria. Di conseguenza, si ha una perdita di energia cinetica e l'urto è detto anelastico. Nella prova sperimentale della gara nazionale delle *Olimpiadi di Fisica 2018* è stato proposto ai partecipanti lo studio di un urto anelastico di un corpo in caduta libera con l'uso di un catetometro artigianale; in particolare, l'esperienza prevedeva la misura del coefficiente di restituzione (ovvero la frazione di energia rimasta dopo l'urto) e della durata dell'urto (ovvero il tempo in cui la pallina scambia quantità di moto con il suolo) [1,2]. L'esperimento ha fornito l'idea di base per uno degli incontri realizzati nell'Anno Scolastico 2020-2021 nell'ambito di "*Lab2Go@Home*", la versione *on-line,* proposta in seguito all'emergenza COVID-19, del progetto Lab2Go. Lab2Go è un'iniziativa di Percorsi per le Competenze Trasversali e per l'Orientamento (PCTO), ex- Alternanza Scuola Lavoro, nata dalla collaborazione tra Sapienza Università di Roma e l'Istituto Nazionale di Fisica Nucleare (INFN), sezione Roma, ma diffusa su gran parte del territorio nazionale grazie alla collaborazione con altre sezioni INFN e supportata dal Piano Lauree Scientifiche (PLS), il cui fine è la diffusione della pratica laboratoriale nelle scuole [3-8]. Lab2Go@Home è l'espressione di Lab2Go che tiene conto dell'emergenza COVID-19 e dell'impatto che essa ha avuto sull'insegnamento sia scolastico che universitario, rendendo necessario il ricorso alla didattica a distanza. Senza dubbio, i disagi legati alla didattica a distanza hanno riguardato in particolar modo le materie scientifiche a causa dell'impossibilità per docenti e studenti di recarsi in laboratorio. Al fine di garantire il prosieguo della pratica laboratoriale anche in questo difficile periodo, si è reso necessario proporre agli studenti esperimenti che potessero essere svolti anche a casa con l'ausilio di materiale a basso costo e facilmente reperibile [9]. L'impegno del progetto Lab2Go in tale direzione è stato di grande rilievo, proponendo alle scuole partecipanti numerose attività volte a promuovere la pratica sperimentale anche senza la possibilità di frequentare i laboratori scolastici [10]. Durante l'incontro del 4 marzo 2021, dedicato agli urti anelastici, è stato mostrato a studenti e docenti delle scuole come costruire un catetometro artigianale e utilizzare tale strumento per misurare il coefficiente di restituzione e il tempo di durata dell'urto come proposti nella gara nazionale delle *Olimpiadi di Fisica 2018*. In aggiunta all'idea originale delle Olimpiadi di Fisica 2018 sono stati mostrati durante l'incontro e proponiamo in questo articolo:
- la ricostruzione dell'andamento esponenziale dell'altezza dopo ogni urto;
- la misura del coefficiente di restituzione con l'utilizzo dell'app *PhyPhox* e della scheda *Arduino*.

Poiché il metodo di misura della durata dell'urto e del coefficiente di restituzione con l'uso del catetometro sono stati già ampiamente discussi, si rimanda alla bibliografia per maggiori dettagli su questa parte; concentrandoci, invece, sull'utilizzo di *PhyPhox* e *Arduino*. In particolare, la misura della durata dell'urto, non è discussa in questo articolo, invece, la misura del coefficiente di restituzione col catetometro è mostrata solo per effettuare il confronto con la stessa misura effettuata con *PhyPhox* e *Arduino*.

1. **Procedura sperimentale per la misura del coefficiente di restituzione**
   Per l'esecuzione dell'esperimento è stato costruito il catetometro artigianale in legno riportato in figura 1 ed è stata utilizzata una pallina da ping-pong, di diametro *d*=(40,00±0,05) mm (misurato con un calibro a nonio ventesimale), lasciata cadere da un'altezza iniziale misurata col catetometro $h_0$=(50,0±0,1) cm.

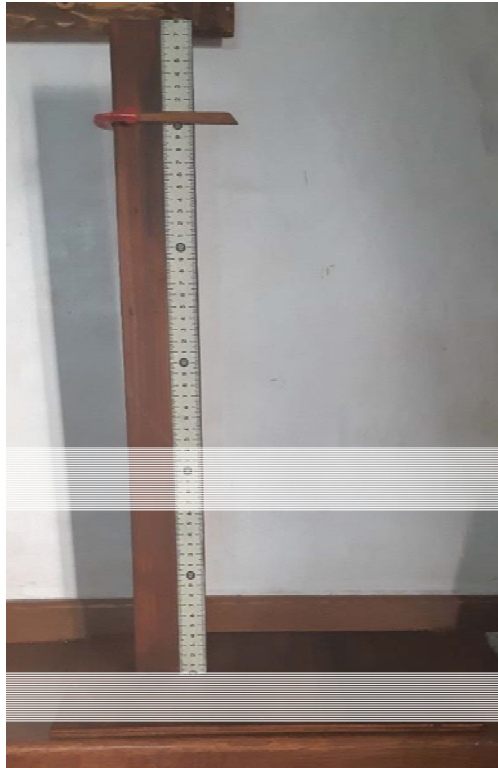
Figura 1: Catetometro artigianale costruito dal borsista Lab2Go Fausto Casaburo.

Ogni misura è stata ripetuta N=3 volte e ne sono stati calcolati il valor medio e la deviazione standard. Detti *i* il numero dell'i-esima collisione e *k<0* un parametro caratteristico dello specifico urto che dipende dai corpi che collidono, l'andamento dell'altezza è descritto dalla funzione:

$$h(i) = h_0 e^{ki} \tag{1}$$

Per linearizzare tale andamento, è stato calcolato il logaritmo di entrambi i membri dell'equazione 1 ed è stata effettuata un'interpolazione dei dati utilizzando il *software* di analisi *ROOT* [11] con la funzione:

$$ln\left(\frac{h}{1\text{m}}\right) = ki + q \tag{2}$$

$dove\ q = ln\left(\frac{h_0}{1\text{m}}\right)$.

### 1.1 Uso del catetometro

Grazie all'utilizzo del catetometro e di uno *smartphone* per visualizzare l'urto in *slow-motion*, sono state misurate le altezze $h_1$, $h_2$, $h_3$ e $h_4$ raggiunte dalla pallina rispettivamente dopo il primo, secondo, terzo e quarto urto (Fig. 2).

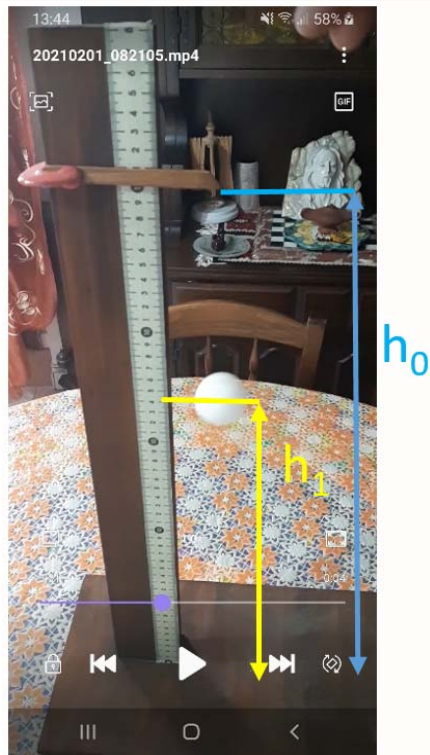

Figura 2: Esempio di misura dell'altezza dopo il primo urto, effettuata con l'utilizzo del catetometro.

Dalle misure di $h_1$, $h_2$, $h_3$ e $h_4$ è stato calcolato il coefficiente di restituzione dopo ogni urto, definito come:

$$e_{h_i} = \frac{h_{i+1}}{h_i} \qquad (3)$$

essendo *i≥0* e da tali valori, assumendo che il coefficiente di restituzione sia costante durante gli urti, è stato calcolato il coefficiente di restituzione medio <*e$_h$*>.

### 1.2 Uso dell'app Phyphox

Nella seconda parte dell'esperienza, è stata utilizzata l'app per *smartphone PhyPhox* [12]; tale app è stata sviluppata dal politecnico di Aachen e tradotta in italiano da studenti di scuola secondaria di II grado durante un progetto ex- *Alternanza Scuola Lavoro* promosso da Sapienza Università di Roma sotto la guida del prof. Giovanni Organtini del Dipartimento di Fisica. Essa sfrutta i sensori dello *smartphone* per effettuare numerosi esperimenti di fisica. Tra i possibili esperimenti, *PhyPhox* presenta "*Collisione (an)elastica*" che permette di misurare il coefficiente di restituzione di un urto. Per questa funzione, l'app fa uso del microfono dello *smartphone* per "sentire" il rumore prodotto dalla pallina durante due urti successivi e ne misura l'intervallo di tempo *Δt*; il tempo misurato è dunque il tempo tra un urto, il raggiungimento di una nuova altezza $h_i$ e l'urto successivo e, poiché trascurando gli attriti i tempi di salita e discesa sono uguali, il tempo di caduta dalla generica altezza $h_i$ è *Δt$_i$/2*. Da tale valore, *PhyPhox* ne ricostruisce l'altezza dalla legge del moto di un corpo in caduta libera, per effetto dell'accelerazione di gravità *g*, con velocità e posizione iniziali nulle:

$$h_i = \frac{1}{2} g \left(\frac{\Delta t_i}{2}\right)^2 \qquad (4)$$

Dai valori di altezza ricavati dopo ogni urto, *PhyPhox* calcola, inoltre, il valore di coefficiente di restituzione (eq. 3) e il valore di energia rimasta ad ogni rimbalzo. Dai valori di coefficiente di restituzione a ogni collisione, ne calcola il valor medio e ricostruisce l'altezza iniziale:

$$h_0 = \frac{h_1}{\langle e_h \rangle} \qquad (5)$$

Come si vede in Figura 3, il risultato ottenuto per $h_0$ con *PhyPhox* è coerente con il valore nominale $h_0$= (50,0 ± 0,1) cm.

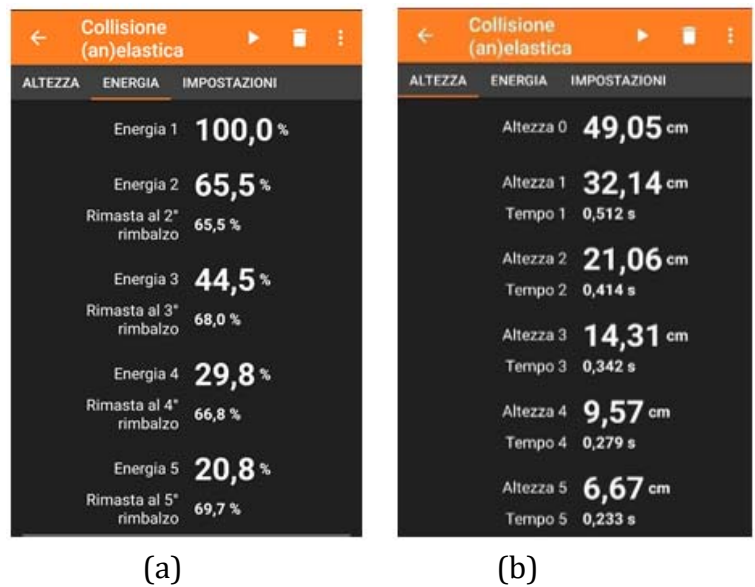

(a)      (b)

Figura 3: Esempio di misura di coefficiente di restituzione (a) e di altezza iniziale (b) con *PhyPhox.*

Per permettere allo *smartphone* di sentire il suono prodotto dalla collisione, senza però rischiare di romperlo, consigliamo di tenerlo a una distanza di circa 10-20 cm dal punto di collisione.

### 1.3 Uso di Arduino

Nell'ultima parte dell'esperimento, abbiamo riprodotto con *Arduino* [13-16] quanto fatto da *PhyPhox*. A partire dall'a.s. 2021/21, infatti, *Arduino* è stato ampiamente utilizzato nell'ambito del progetto Lab2Go [17] permettendo, grazie ai numerosi sensori e al suo basso costo, di effettuare numerosi esperimenti e di sviluppare negli studenti competenze trasversali quali il coding. Per questa misura, è stata utilizzata una scheda *Arduino Uno R3*, un sensore di suono *KY-037* [18] e cavi di collegamento *Dupont* (Fig. 4).

Il sensore *KY-037* è dotato di un microfono e di un regolatore di soglia di silenzio e si può collegare ad *Arduino* sia tramite pin analogico che digitale. Nel nostro caso, il sensore è stato collegato a un ingresso analogico. Il sensore fornisce una tensione compresa tra 0V e la tensione di alimentazione, proporzionale al livello d'intensità sonora rilevata. Gli ingressi analogici di *Arduino* sono connessi a un *Analog to Digital Converter* (ADC) a 10bit, attraverso il quale è possibile misurare tale tensione e, di conseguenza, l'intensità del suono.

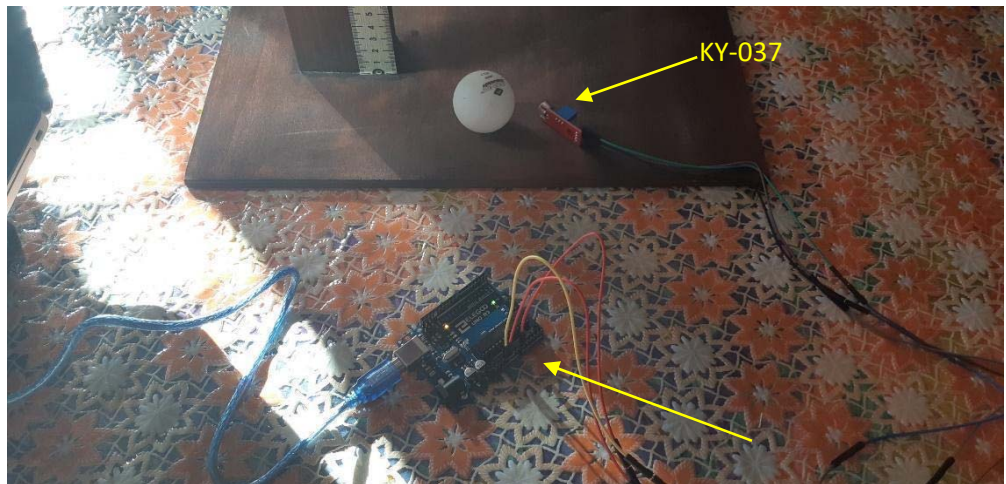

Figura 4: Apparato sperimentale per le misure effettuate con *Arduino*.

La prima operazione è consistita nella realizzazione di un programma che misura il valore dell'intensità sonora ambientale, in modo da stimare la *baseline*, ovvero il valore della soglia sotto la quale si può assumere il "silenzio". Successivamente, utilizzando tale valore, è stato scritto un programma per la misura dei tempi in cui si rileva un suono di intensità superiore alla soglia; nel corso dell'esperimento, in tal modo, un evento nel quale il valore di intensità sonora registrata è superiore alla soglia impostata corrisponde a un urto. È importante, pertanto, che l'esperimento sia svolto in assenza di altre sorgenti di rumore che possano superare la soglia di silenzio. Dai tempi misurati a ogni urto sono stati calcolati il coefficiente di restituzione (Eq. 3) e le altezze sia dopo ogni urto (Eq. 4), sia iniziale (Eq. 5).

## 2. Risultati

Nella figura seguente sono riportati gli andamenti di altezza ricostruiti in funzione della collisione, rispettivamente, dai dati ottenuti con l'uso del catetometro (Fig. 5a), di *PhyPhox* (Fig. 5b) e di *Arduino* (Fig. 5c). I grafici riportano il logaritmo dei valori misurati e il fit lineare (Eq. 2).

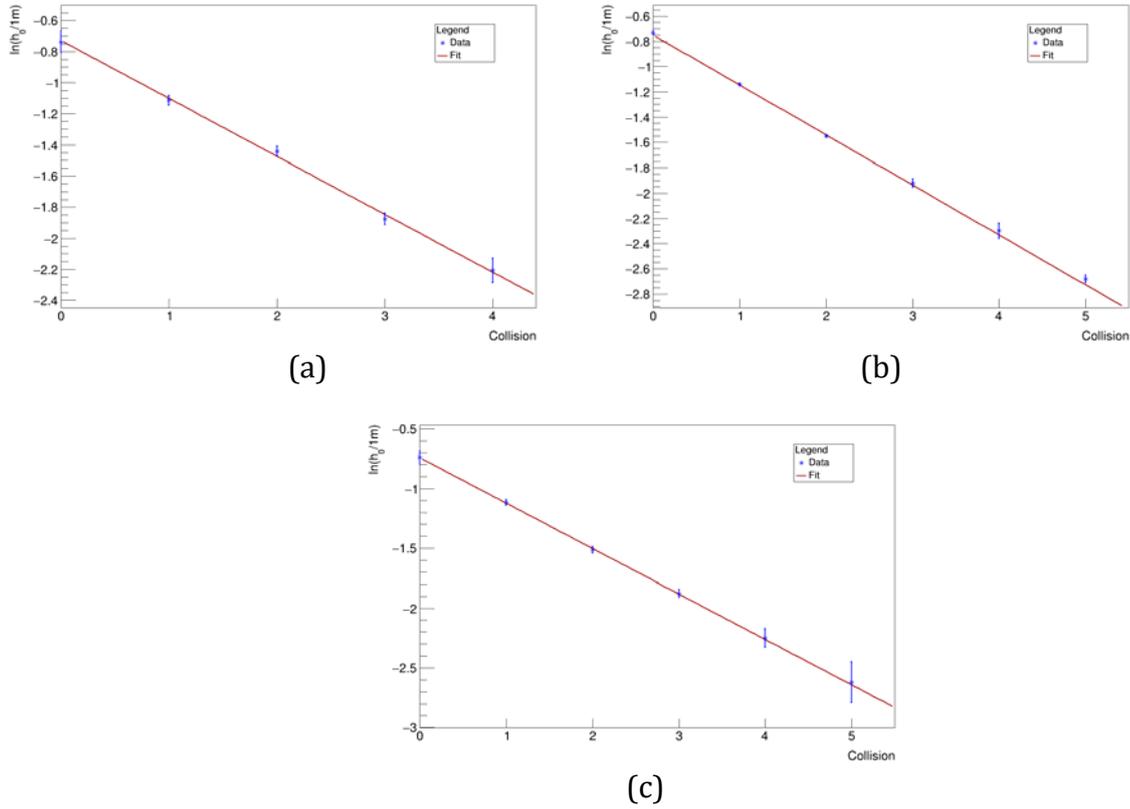

Figura 5: Ricostruzione dell'altezza iniziale (a) con l'uso del catetometro, (b) con l'uso di PhyPhox, (c) con l'uso di Arduino.

Il coefficiente di restituzione (Eq. 3) e i risultati dei fit ottenuti (Fig. 5) sono riportati in tabella 1.

Tabella 1: Risultati ottenuti del coefficiente di restituzione $e_h$ e dei parametri dei fit..

| Strumento | $e_h$ | $ln\left(\frac{h_0}{1m}\right)$ | $k$ |
|---|---|---|---|
| Catetometro | $(6,9 \pm 0,5) \cdot 10^{-1}$ | $(-7,3 \pm 0,4) \cdot 10^{-1}$ | $(-3,7 \pm 0,2) \cdot 10^{-1}$ |
| *PhyPhox* | $(6,8 \pm 0,2) \cdot 10^{-1}$ | $(-7,5 \pm 0,1) \cdot 10^{-1}$ | $(-3,95 \pm 0,06) \cdot 10^{-1}$ |
| *Arduino* | $(6,9 \pm 0,4) \cdot 10^{-1}$ | $(-7,4 \pm 0,3) \cdot 10^{-1}$ | $(-3,8 \pm 0,1) \cdot 10^{-1}$ |

Per quantificare l'accordo tra i risultati riportati nella tabella precedente, è stato innanzitutto effettuato un test del $\chi^2$. In questo caso, assumiamo che il modello (ipotesi zero) sia la funzione costante

$$y = cost. \tag{6}$$

ossia che i punti sperimentali ottenuti dalle tre procedure di misura siano fluttuazioni casuali rispetto al valore medio rappresentato dal risultato del fit costante ottenuto. I grafici dei tre fit effettuati di $e_h$, $ln\left(\frac{h_0}{1m}\right)$, $k$ sono riportati rispettivamente nella figura 6.

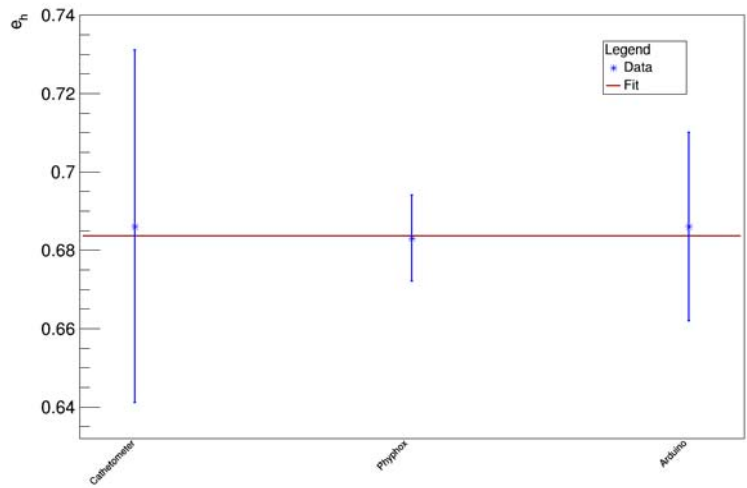

(a)

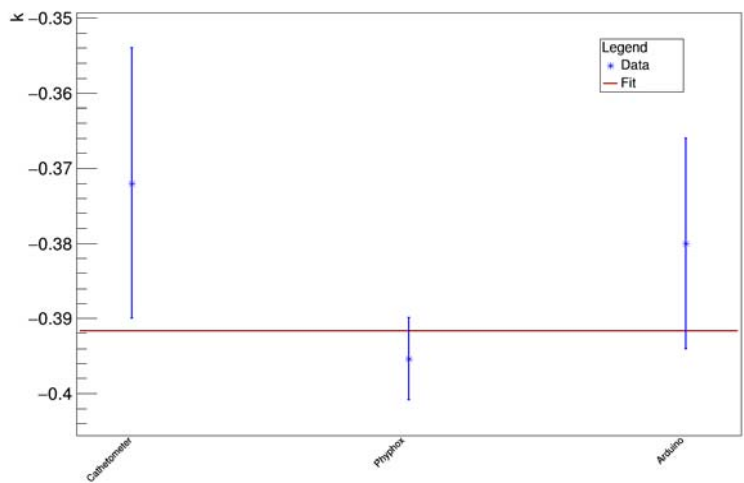

(b)

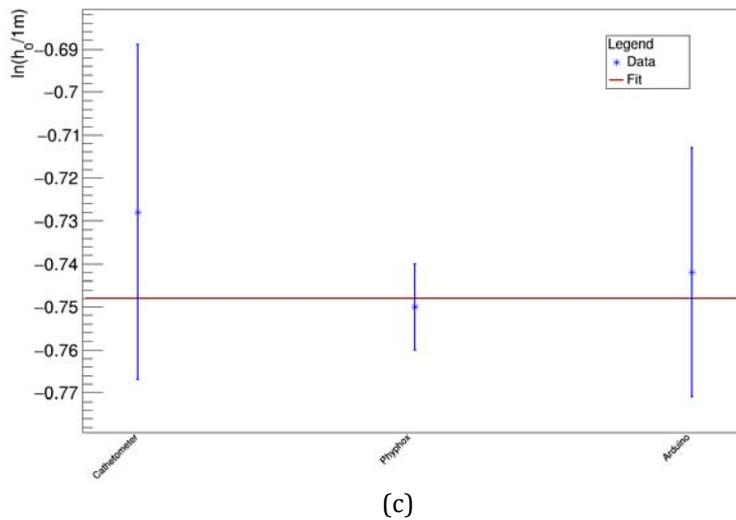

(c)

Figura 6: Fit alle misure (a) del coefficiente di restituzione $e_h$, (b) del parametro k (c) del parametro relativo all'altezza iniziale $ln\left(\frac{h_0}{1\text{m}}\right)$.

Dal risultato del fit e dal valore del $\frac{\chi^2}{Ndf}$ ottenuti (in approssimazione di calcolo del $\chi^2$ di ogni parametro preso singolarmente), dove Ndf è il numero di gradi di libertà (Ndf=2 in questo caso), è stato calcolato il *p-value* [19,20]. Infine, dal valore del *p-value* ottenuto, calcolando la Cumulative Distribution Function (CDF) della funzione gaussiana [21] è stato determinato il numero di $\sigma$ di accordo, $N_\sigma$, tra i risultati [22].

I risultati di valor medio e $\frac{\chi^2}{Ndf}$ ottenuti dai fit (Fig. 6a, Fig, 6b, Fig. 6c), di *p-value* e $N_\sigma$ sono riportati in tabella. 2.

Tabella 2: Accordo tra le misure effettuate con catetometro, PhyPhox e Arduino..

|  | $e_h$ | $ln\left(\frac{h_0}{1m}\right)$ | $k$ |
| --- | --- | --- | --- |
| Mean | $(6,8 \pm 0,1) \cdot 10^{-1}$ | $(-7,48 \pm 0,09) \cdot 10^{-1}$ | $(-3,92 \pm 0,05) \cdot 10^{-1}$ |
| $\frac{\chi^2}{Ndf}$ | $\sim \frac{0,02}{2}$ | $\sim \frac{0,35}{2}$ | $\sim \frac{2,35}{2}$ |
| *p-value* | $\sim 1$ | $\sim 8 \cdot 10^{-1}$ | $\sim 3 \cdot 10^{-1}$ |
| $N_\sigma$ | $\sim 1 \cdot 10^{-2}$ | $\sim 2 \cdot 10^{-1}$ | $\sim 1$ |

Dai valori riportati in Tab.2, si osserva che i valori di *p-value* sono maggiori di 0,05 (per cui il test è verificato al 95% CL) e i risultati sono compatibili tra loro entro $\sim 1\sigma$.

**Conclusioni**

La pratica laboratoriale è, purtroppo, spesso trascurata nelle scuole secondarie di II grado, nonostante essa abbia un ruolo fondamentale nell'insegnamento e nell'apprendimento delle discipline scientifiche. Lo scopo di *Lab2Go* è da sempre stato avvicinare gli studenti alla scienza. In questo periodo di pandemia, per far fronte alle difficoltà nel portare i ragazzi in laboratorio, sono stati proposti incontri *on-line* in cui sono stati mostrati esperimenti riproducibili, anche a casa, con strumenti facilmente reperibili, con l'uso dell'app *PhyPhox* o della scheda programmabile *Arduino*. L'esperienza sarà utile anche una volta finita l'emergenza COVID-19 nei laboratori scolastici meno attrezzati, o qualora si vogliano proporre dimostrazioni in aula senza spostare le classi in laboratorio. L'esperimento riportato in questo articolo è stato mostrato durante uno degli incontri *on-line* di *Lab2Go* agli studenti e ai docenti delle scuole partecipanti al progetto, suscitando molto interesse, anche in considerazione dell'argomento, gli urti, che è presente nei programmi curriculari, ma è spesso trattato facendo riferimento al solo caso ideale (urto elastico). A tal proposito, le testimonianze raccolte tra gli studenti hanno mostrato che l'esperienza proposta è stata molto utile per approfondire le conoscenze teoriche sugli urti acquisite a scuola. Il coinvolgimento è stato tale da spingere alcuni di loro a ripetere l'esperimento autonomamente a casa. Inoltre, in generale, gli studenti hanno riferito di essere soddisfatti dal progetto Lab2Go che, nonostante le difficoltà dovute all'impossibilità di recarsi in laboratorio a causa dell'emergenza COVID-19, ha saputo suscitare grande interesse in loro verso la fisica sperimentale. Vogliamo concludere ricordando che le caratteristiche dell'urto dipendono dai corpi in contatto e che quello qui riportato (urto di una pallina da ping-pong sulla base in legno del catetometro) è solo un esempio; l'invito agli studenti è quello, quindi, di ripetere l'esperimento proposto variando i corpi che collidono, provando diversi materiali e altre tipologie di palline o di superfici (legno, plastica, metallo, tessuto). Infatti, al di là dei risultati numerici riportati e dell'accordo ottenuto, lo scopo di *Lab2Go* è far avvicinare i ragazzi alla scienza sperimentale, fornire indicazioni su come effettuare una misura, interpretare e gestirne gli errori. Ci auguriamo, quindi, che questo lavoro possa essere uno stimolo per i lettori di *"La Fisica nella Scuola"* a ripetere l'esperienza e proporne ulteriori varianti. Infine, seppur non in ordine d'importanza, auspichiamo che i docenti che leggeranno questo articolo vorranno proporre alla loro scuola, dal prossimo anno scolastico, la partecipazione al progetto *Lab2Go.*